\documentstyle[12pt]{article}

\newcommand{\be}{\begin{equation}}
\newcommand{\ee}{\end{equation}}
\newcommand{\mpl}{ {M_{\rm pl}}} 
\newcommand{\mbh}{ {M_{\rm bh}}} 
\newcommand{\mo}{ {M_{0 \rm bh}}} 
\newcommand{\cct}{ {T_{\rm vac}}} 
\newcommand\pp{\parshape 2 0.0truecm 14.25truecm 1.25truecm 13.0truecm}

\begin{document}
\baselineskip=20pt 

\centerline{\bf POSSIBLE EFFECTS OF A COSMOLOGICAL CONSTANT} 
\centerline{\bf	ON BLACK HOLE EVOLUTION} 

\bigskip 
\centerline{\bf Fred C. Adams and Manasse Mbonye} 
\bigskip 
\centerline{Physics Dept., University of Michigan, Ann Arbor, MI 48109}
\bigskip 
\centerline{\bf Gregory Laughlin} 
\bigskip 
\centerline{Astronomy Dept., University of California, Berkeley, CA 94720}
\bigskip 
\bigskip 
\centerline{\it 22 January 1999} 
\bigskip 

\begin{abstract} 
We explore possible effects of vacuum energy on the evolution of black
holes.  If the universe contains a cosmological constant, and if black 
holes can absorb energy from the vacuum, then black hole evaporation 
could be greatly suppressed.  For the magnitude of the cosmological 
constant suggested by current observations, black holes larger than 
$\sim 4 \times 10^{24}$ g would accrete energy rather than evaporate.  
In this scenario, all stellar and supermassive black holes would grow 
with time until they reach a maximum mass scale of 
$\sim 6 \times 10^{55}$ g, comparable to the mass contained 
within the present day cosmological horizon. 
\end{abstract}

\bigskip 


keywords: black holes, cosmological constant

{\vfil\eject}
\bigskip 
\medskip 

The past several years have presented us with two important and
intriguing developments concerning the nature of our universe: 

[A] Observations of distant supernovae [1] strongly suggest that the
Hubble expansion departs from that expected for a purely matter
dominated cosmology. The leading explanation for this departure is a
cosmological constant contribution to the energy density of roughly
half the critical density, i.e., $\rho_V$ $\approx$ $\rho_{cr}/2$
$\approx$ $10^{-29}$ g/cm$^3$ $h^2$ $\approx$ (0.003 eV)$^4$ $h^2$,
where $h$ is the present day Hubble constant in units of 100 
km s$^{-1}$ Mpc$^{-1}$ so that $0.4 < h < 1$. The corresponding 
energy scale of the vacuum is thus $\cct$ = 0.003 eV $h^{1/2}$. 

[B] The observational evidence for black holes has passed a 
threshold of firmness so that black holes can now be considered as
``discovered''. This observational evidence can be found in three
different settings: the three million solar mass black hole in the
center of our galaxy [2], supermassive black holes in the centers of
external galaxies [3], and stellar mass black holes within our galaxy
[4]. Thus far, however, no evidence has been found for smaller black 
holes [5], which presumably have a primordial origin. 

Given the existence of both black holes and a cosmological constant,
an interesting physical process can potentially occur: The black holes
can accrete energy from the vacuum and grow larger with time [6]. The
usual conceptual description of a cosmological constant is that
seemingly empty space is not really empty, but rather is continually
seething with virtual particles, which must contribute a net positive
energy density.  Within this picture, the virtual particles can be
accreted by black holes. Given the (almost) one-way nature of a black
hole's event horizon, more energy will enter the black hole than will
be released and the black hole can gain energy and thereby grow
larger.  In this letter, using the explicit assumption that this
accretion process is viable, we explore the possible effects of a
non-vanishing cosmological constant on the future evolution of black
holes.

Building on earlier work [7, 8], Mallett [6] performed the
relativistic calculation of an evaporating black hole embedded within
a background space-time endowed with a cosmological constant.  The
original motivation was to determine the effects of the vacuum energy
on the evaporation of black holes during the inflationary epoch, but
the results apply to the present case as well.  For a black hole
radiating into a background universe with a vacuum energy
contribution, the line element can be written in advanced time
coordinates in the form 
\be 
ds^2 = - \bigl[ 1 - {2 G M(v) \over r} - \chi^2 r^2 \bigl] dv^2 
+ 2 dv dr + r^2 d \Omega^2 \, . 
\ee
Unless explicitly stated otherwise, we work in units with 
$\hbar = 1 = c$ and hence $G = 1/\mpl^2$. The parameter $\chi$ 
sets the magnitude of the cosmological constant and is defined 
by the relation 
\be 
\chi = \Bigl({2 \pi^3 \over 45}\Bigr)^{1/2} {\cct^2 \over \mpl} \, , 
\ee 
and where $\cct$ is the effective temperature scale of the 
cosmological vacuum energy ($\cct$ $\approx$ 0.003 eV $\approx$ 
34 K for the presently suspected cosmological constant). 

Following Mallett [6], we can write the effective luminosity of a 
black hole living within this space-time in the phenomenological form 
\be 
L = 4 \pi (r^{-}_{AH})^2 \bigl\{ T_H^4 - \cct^4 \bigr\} \, , 
\label{eq:lum} 
\ee
where $r^{-}_{AH}$ is the inner apparent horizon and $T_H$ is 
the usual Hawking temperature (this equation represents a conjecture, 
rather than a rigorous derivation, so we ignore dimensionless constants 
of order unity). The first term represents the Hawking radiation 
flux [9] flowing outward from the black hole, whereas the second 
term represents an inward accretion of energy from the vacuum [6]. 
The apparent horizons are determined by the condition 
\be 
1 - {2 G M(v) \over r} - \chi^2 r^2 = 0 \, . 
\label{eq:ahorizon}
\ee
For black holes of astrophysical interest, the mass lies in the 
range $1 M_\odot < M_{\rm bh} < 10^{10} M_\odot$. For these hole 
masses and the suggested value of $\cct$, the inner apparent horizon 
$r^{-}_{AH}$ is close to the Schwarzschild radius $r_S$ = $2 G M$ 
and the Hawking temperature is close to the usual result 
$T_H = 1/8 \pi G M$ [9, 10, 11]. 

The inward accretion flow implied by equation (\ref{eq:lum}) can be
motivated by a simple conceptual argument analogous to that often used
for the Hawking effect. Consider the volume of space located just
outside the black hole horizon. The volume is filled with virtual
particles with characteristic energy $\cct$ and wavelength $\lambda =
1/\cct$.  If a given particle has a large uncertainty $\Delta r$ in
its radial position, then it could happen to lie within the black hole
horizon and can be accreted directly.  So let's suppose that the
particle does not have a large uncertainty in its radial position.  
In particular, it must be localized so that $\Delta r$ $<$ 
$\lambda$. Then the uncertainty principle implies that the radial
momentum $p_r$ obeys the ordering $p_r \approx \Delta p_r$ $>$
$1/\Delta r > 1 / \lambda = \cct$.  On average, the momenta for half
of such particles will be directed radially inward.  This inward
momentum implies a net accretion flux of $\sim n p_r/2$, where 
$n \approx \cct^3$ is the number density of the particles.  With this
flux, and with the effective area $4 \pi (r^{-}_{AH})^2$ of the black
hole horizon, the net accretion rate becomes 
\be 
{\dot M} = 4 \pi (r^{-}_{AH})^2 \cct^4 \, 
\approx 3 \, {\rm g} \, {\rm yr}^{-1} h^2 
\Bigl( {M_{\rm bh} \over 1 M_\odot} \Bigr)^2 \, , 
\ee 
in accordance with equation (\ref{eq:lum}). 

For comparison, recall that a de Sitter space with no black holes
produces a thermal radiation bath with an effective temperature of
$T_{\rm dS} \sim \chi$ $\sim \cct^2/\mpl$ [10, 12]. This radiation 
results from a Hawking-like effect in which the cosmological horizon 
at $r \approx 1/\chi$ emits nearly thermal radiation into the 
universe. Black holes will also accrete this energy [6], and hence
equation (\ref{eq:lum}) should contain an additional term $\propto
\chi^4$.  This radiation is less energetic than the vacuum energy
scale $\cct$ by nearly 31 orders of magnitude, however, and its
contribution to black hole accretion is negligible in this present
context.

For completeness, we also note that black holes absorb energy
from the cosmic background radiation field. At the present epoch, 
these microwave background photons have an effective temperature of
$T_{\rm cmb}$ = 2.74 K $\approx$ 0.00024 eV, about ten times smaller
than $\cct$. As a result, this additional contribution to the 
accretion flux is approximately $10^4$ times smaller than that 
due to the $\cct^4$ term.  The accretion of energy from the cosmic 
background will become increasingly less important as the universe 
expands and the photons redshift. In contrast, the energy density 
of the vacuum remains constant. 

This problem contains an important critical mass scale $M_C$.
Sufficiently small black holes will experience Hawking evaporation 
and are relatively unaffected by the presence of the cosmological 
constant. For large black holes, however, the Hawking temperature
$T_H$ is less than $\cct$ and such black holes can accrete energy from
the vacuum rather than evaporate.  The critical mass scale $M_C$,
obtained by setting $T_H$ = $\cct$, has a value of 
\be 
M_C = {1 \over 8 \pi G \cct} \approx 4 \times 10^{24} {\rm g} 
\approx 2 \times 10^{-9} M_\odot \, . 
\ee
This mass scale is about the same as that of Titania, the largest moon
in the Uranian system [13]. Thus, all black holes more massive than
Titania will accrete energy rather than evaporate.  The Schwarzschild
radius of such a critical mass black hole would be rather small, only 
about 6 microns.

A second characteristic mass scale exists.  For sufficiently large 
black hole masses, the cubic equation (\ref{eq:ahorizon}) has no real 
positive solutions and the apparent horizons disappear. This condition
defines a second critical mass $M_\ast$ given by 
\be 
M_\ast = {1 \over 3 \sqrt{3} \chi G} \approx 6 \times 10^{55} 
{\rm g} \approx 3 \times 10^{22} M_\odot \, , 
\ee 
where we have used the presumed magnitude of the cosmological 
constant.  This critical mass scale, with the mass equivalent of 
$3 \times 10^{79}$ protons, is thus roughly comparable to the mass 
contained within the present-day cosmological horizon. 

Given enough time, a black hole can accrete energy from the vacuum
until it reaches the second critical mass scale $M_\ast$. The total
accretion time $\tau$ is defined to be the time required for a black
hole with initial mass $\mo$ (above the minimum mass threshold $M_C$) 
to accrete enough energy to shed its horizons.  This accretion time 
is given by 
\be 
\tau = {1 \over 4 \pi \cct^4} \int_{\mo}^{M_\ast} 
{dM \over r^2 (M) } \, = \, \Bigl( {2 \pi \over 15} \Bigr)^{1/2} 
{\mpl \over 8 \cct^2} {(1-q)^2 \over q} \, , 
\ee
where $\mo$ is the starting mass of the black hole, 
$r(M)$ is the inner apparent horizon as defined by equation 
(\ref{eq:ahorizon}), and $q$ is the root of the equation 
$q^3 - 3 q + 2 \mo / M_\ast = 0$.  For all known black holes, 
$\mo/M_\ast \ll 1$, $q \approx 2 \mo/3 M_\ast$, and the time 
scale becomes 
\be 
\tau \approx 0.02 (\mpl/\cct)^4 \mo^{-1} \approx 5 \times 10^{31} 
{\rm yr} \, h^{-2} \, \Bigl( {\mo \over 1 M_\odot} \Bigr)^{-1} \, . 
\ee 

Black holes thus have a large, but strictly finite, dynamical range 
in this scenario.  Relatively large black holes (with initial mass  
$\mo > M_C$) grow to the maximum mass scale $M_\ast$ $\sim$ $10^{56}$
g, whereas smaller black holes (with mass $\mo < M_C$) shrink to the
Planck mass $\mpl$ $\sim 10^{-5}$ g through the Hawking effect. Black
holes are thus confined to a mass range that is ``only'' 61 orders 
of magnitude in extent. 

Even though relatively large black holes (with $M_{\rm bh} > M_C$) 
accrete energy rather than evaporate, they continue to emit  
a flux of ``ordinary'' radiation through the Hawking process. 
The curvature of space-time near the event horizon gives rise 
to a nearly thermal spectrum of photons, neutrinos, and gravitons 
emerging from the hole [14].  With this luminosity $L_H$, the 
black hole thus follows its usual evolutionary track in the 
Hertzsprung-Russell diagram, i.e., 
\be 
L_H = {\sigma_B \over 4 \pi} T_H^2 \, . 
\ee 
With no accretion, the Hawking temperature $T_H$ increases with 
time as the black hole mass decreases; with a net accretion, 
however, the temperature $T_H$ is a decreasing function of time.

\bigskip 

In summary, we have explored the possible consequences of nonzero
vacuum energy on black hole evolution.  In particular, we have
considered a scenario in which the vacuum energy can be accreted in
accordance with equation (\ref{eq:lum}).  If our universe does indeed
contain a substantial fraction of its energy density in the form of a
cosmological constant contribution, then the long term fate and
evolution of black holes can be greatly altered:

[1] All known black holes will never evaporate through the Hawking 
effect. Instead they will continue to grow larger by accretion 
of energy from the vacuum (the cosmological constant energy). 

[2] The only black holes that can ever be observed to evaporate in 
the present day universe must lie within the restricted mass range 
$4 \times 10^{15}$ g $< \mbh <$ $4 \times 10^{24}$ g. Smaller
black holes will have evaporated by the current cosmological epoch,
whereas larger black holes will accrete energy rather than evaporate.
The larger black holes that accrete energy will continue to emit 
photons, neutrinos, and gravitons through the Hawking process, 
but with an ever decreasing temperature. 

[3] In the long term, if black holes continue to accrete energy 
and grow larger, the horizons will vanish as the black hole mass
approaches a critical mass scale $M_\ast$ which is comparable to 
the present day horizon mass scale (the mass equivalent of about 
$3 \times 10^{79}$ protons). 

This evolutionary scenario for black holes rests on the validity 
of equation (\ref{eq:lum}), which ultimately depends on the nature 
of the vacuum energy.  A full understanding of this issue thus 
requires a solution to the cosmological constant problem, which 
remains an open question [15]. This effect greatly changes the 
long term evolution of black holes, however, and could have 
important implications for the long term fate of our universe [16]. 
This present discussion does not address the back reaction, i.e., 
the effects of black hole accretion on the background cosmological 
constant. This issue must be addressed to obtain a full understanding 
of this effect.

\bigskip 

We would like to thank Ron Mallett for many insightful discussions.  
We also thank R. Akhoury, M. Einhorn, and G. Kane for useful criticism 
of the manuscript.  This work was supported by funds from the University
of Michigan.

{\vfil\eject}
\baselineskip=16pt 
{\bf References} 
\medskip 

\medskip\par\pp{[1]}
S. Perlmutter et al., Nature {\bf 391} (1998) 51; 
P. M. Garnavich et al., ApJ {\bf 509} (1998) 74; 
A. G. Reiss et al., ApJ (1999) in press. 

\medskip\par\pp{[2]} 
R. Genzel et al., ApJ {\bf 472} (1996) 153; 
A. M. Ghez et al., ApJ {\bf 509} (1998) 678. 

\medskip\par\pp{[3]} 
J. Kormendy, et al., ApJ {\bf 482} (1997) L139. 

\medskip\par\pp{[4]} 
R. Narayan, D. Barret, and J. E. McClintock, ApJ, 
{\bf 482} (1997) 448. 

\medskip\par\pp{[5]} 
B. J. Carr, ApJ {\bf 206} (1976) 8. 

\medskip\par\pp{[6]} 
R. L. Mallett, Phys. Rev. D {\bf 31} (1985) 416; 
R. L. Mallett, Phys. Rev. D {\bf 33} (1986) 2201; 
R. L. Mallett, Phys. Rev. D {\bf 34} (1986) 1916.  

\medskip\par\pp{[7]} P. Hajicek and W. Israel, Phys. Lett. {\bf 80 A} 
(1980) 9; J. Bardeen, Phys. Rev. Lett. {\bf 46} (1981) 382. 

\medskip\par\pp{[8]} 
J. W. York Jr., in: {\sl Quantum Theory of Gravity: Essays in Honor 
of the Sixtieth Birthday of Bryce DeWitt}, ed. S. Christensen 
(Hilger, Bristol, 1984) p. 135. 

\medskip\par\pp{[9]} 
S. W. Hawking, Comm. Math. Phys. {\bf 43} (1974) 199; 
S. W. Hawking, Nature {\bf 248} (1974) 30. 

\medskip\par\pp{[10]} 
N. D. Birrell and P. C. W. Davies, {\sl Quantum Fields in Curved Space} 
(Cambridge Univ. Press, Cambridge, 1982). 

\medskip\par\pp{[11]} 
K. S. Thorne, R. H. Price, and D. A. MacDonald, {\sl Black Holes: 
The Membrane Paradigm} (Yale Univ. Press, New Haven, 1986). 

\medskip\par\pp{[12]} 
S. A. Fulling, J. Phys. A {\bf 10} (1977) 917. 

\medskip\par\pp{[13]} 
F. H. Shu, The Physical Universe (Univ. Science Books, Mill Valley, 1982). 

\medskip\par\pp{[14]} 
D. N. Page, Phys. Rev. D {\bf 13} (1976) 198; 
D. N. Page, Phys. Rev. D {\bf 14} (1976) 3260.

\medskip\par\pp{[15]} 
S. Weinberg, Rev. Mod. Phys. {\bf 61} (1989) 1. 

\medskip\par\pp{[16]} 
F. C. Adams and G. Laughlin, Rev. Mod. Phys. {\bf 69} (1997) 337. 

\end{document}